\documentclass[12pt,preprint]{aastex}

\usepackage{color}
\definecolor{darkblue}{rgb}{0.0, 0.0, 0.5}

\usepackage{hyperref}
\hypersetup{colorlinks=true,citecolor=darkblue,linkcolor=darkblue}

\slugcomment{ApJ accepted 05 Dec 2012}

\shorttitle{Companion Limits to Fomalhaut}
\shortauthors{Kenworthy et al.}

\begin{document}

\title{Coronagraphic Observations of Fomalhaut \\
at Solar System Scales}

\author{Matthew A. Kenworthy and Tiffany Meshkat}
\affil{Leiden Observatory, Leiden University, P.O. Box 9513, 2300 RA Leiden, The Netherlands}
\author{Sascha P. Quanz}
\affil{Institute for Astronomy, ETH Zurich, Wolfgang-Pauli-Strasse 27, 8093 Zurich, Switzerland}

\author{Julien H. Girard}
\affil{European Southern Observatory, Alonso de Cordova 3107, Vitacura, Cassilla 19001, Santiago, Chile}

\author{Michael R. Meyer}
\affil{Institute for Astronomy, ETH Zurich, Wolfgang-Pauli-Strasse 27, 8093 Zurich, Switzerland}

\and

\author{Markus Kasper}
\affil{European Southern Observatory, Karl Schwarzschild Strasse, 2,
85748 Garching bei Munchen, Germany}

\begin{abstract}

We report on a search for low mass companions within 10 AU of the star
Fomalhaut, using narrow band observations at $4.05\mu m$ obtained with
the Apodizing Phase Plate (APP) coronagraph on the VLT/NaCo. Our
observations place a model dependent upper mass limit of $12-20M_{jup}$
from 4 to 10 AU, covering the semi-major axis search space between
interferometric imaging measurements and other direct imaging
non-detections. These observations rule out models where the large
semi-major axis for the putative candidate companion Fomalhaut b is
explained by dynamical scattering from a more massive companion in the
inner stellar system, where such giant planets are thought to form.

\end{abstract}


\section{Introduction}

The formation and distribution of the planets in our Solar system is
closely entwined with the early evolution of its debris disk
\citep{Lissauer93} which, despite many different lines of evidence, do
not produce a clear picture.  By studying debris disks with a range of
stellar ages, however, an evolutionary picture can be formed \citep[see
the reviews of][]{Wyatt07,Meyer07}.  The morphology of dust in debris
disks around nearby stars is thought to be a signpost of planet
formation, leading to an intense study of the nearest debris disk
systems to look for the bodies that are sculpting these structures. Deep
imaging surveys therefore have focused on the closest systems where
spatial resolution provides the greatest detail. Fomalhaut is a nearby
($d=7.7pc$) A3V star with an estimated age of $440\pm40$ Myr
\citep{Mamajek12}. It has an inclined eccentric debris disk first
resolved by \citet{Holland98} in the sub-mm and subsequently imaged by
the Hubble Space Telescope by \citet{Kalas05}. The sharp inner edge of
the resolved ring and its eccentricity implies the gravitational
presence of a planet \citep{Quillen06} for which a candidate was imaged
by \citet{Kalas08} at optical wavelengths. 

Observations taken with the Herschel Space Observatory from $70 \mu m$
to $500 \mu m$ \citep{Acke12} show that there is an estimated 110 Earth
masss cometary reservoir supplying the dusty grains that are directly
detected at these wavelengths. High angular resolution observations at
350 GHz taken with ALMA resolves the debris ring on one side of
Fomalhaut \citep{Boley12}. The structure of the ring implies that the
outer edge of the parent body ring is consistent with being as sharply
truncated as the inner edge. The most likely explanation is the presence
of unseen shepherding planetary bodies on either side of the ring.

We report on the search for a substellar companion to Fomalhaut down to
3 AU of the star. We are motivated by the presence and location of the
object labelled Fomalhaut b, which resides approximately 120 AU in
projection at the inner edge of the debris ring. The nature of Fomalhaut
b is uncertain. The initial detection of reflected light in
\citet{Kalas08} was made at visible wavelengths, but subsequent
observations do not detect the expected thermal emission from a massive
gas giant \citep{Marengo09,Janson12}. The continuing existence of the
dust ring implies a mass of less than three Jupiter masses, a predicton
that is confirmed by a Spitzer non-detection of Fomalhaut b which limits
its mass to $1\,M_{jup}$ \citep{Janson12}.  The distance of Fomalhaut b
from its parent star means that the reflected light detection from the
atmospheric surface of a Jupiter mass planet is not possible. Instead,
the blue colours of the object imply we are seeing starlight scattered
from a cloud of dust which is either organised in a ring system about an
unseen central object or a recently formed cloud of ejecta from a
collision of planetesimals. A third alternative is that we are seeing a
resonant clump of dust formed by unseen peturbation in the debris ring
system.

It is a challenge to current planet formation theories to explain the
presence of a massive object in an orbit at that distance from its
central star, leading to the hypothesis that such planets are captured
from another star during the early stages of stellar cluster evolution
\citep{Parker12}, or that the object formed much closer to the parent
star and was then dynamically ejected out to 120 AU through
gravitational interactions with one or more massive planets in the inner
Fomalhaut system \citep{Chiang09}.  Additionally it is interesting in
its own right to look for giant planets around early type stars, given
that the first gas giant exoplanet detections have been around the
early-type stars $\beta$ Pictoris \citep{Lagrange09,Lagrange10,Quanz10}
and HR 8799 \citep{Marois08,Marois10,Skemer12}, and to this end direct
imaging surveys are producing statistically significant samples for
analysis and interpretation \citep{Ehrenreich10,Janson11,Vigan12}.

In Section \ref{obs} we describe the observations and data reduction
carried out with NaCo at the VLT, in Section \ref{resan} we carry out an
analysis and interpretation of the presented data in terms of limits to
faint companions of Fomalhaut. In Section \ref{discus}, we compare our
search with other companion searches around Fomalhaut and how future
observations with the APP will be improved, and in Section \ref{concl}
we present our conclusions.

\section{Observations and Data Reduction\label{obs}}

Data were obtained using NaCo \citep{Lenzen03,Rousset03} on UT4 at
Paranal in July and August 2011 (see Table \ref{tabobs}). The
conditions were photometric with no cloud cover. NaCo is run with the
target, Fomalhaut, used as a natural guide star and the visible band
wavefront sensor.

\begin{deluxetable}{llll}
\tablecaption{Summary of Observations of Fomalhaut in Pupil Tracking
Mode\label{tabobs}}
\tablewidth{0pt}
\tablehead{
    \colhead{Parameter} & \colhead{Hemisphere 1} & \colhead{Hemisphere 2} & \colhead{Hemisphere 3}
}

\startdata
UT date  & 2011-07-19   & 2011-08-09 & 2011-08-07 \\
UT start & 07:27:43     & 06:08:23   & 06:04:17   \\
UT end   & 08:42:30     & 07:25:20   & 07:15:04   \\
NDIT $\times$ DIT & $200\,\times\,0.23$ sec & $200\,\times\,0.23$ sec  &
$200\,\times\,0.23$ sec    \\
Number of Cubes & 75 & 78 & 72 \\
Parallactic Angle start & -50.2$^\circ$ & -46.1$^\circ$  &  -58.9$^\circ$ \\
Parallactic Angle end & 69.3$^\circ$ &  71.7$^\circ$  &  61.8$^\circ$  \\
Airmass & 1.005 to 1.023 & 1.005 to 1.014 & 1.005 to 1.028 \\
Typical DIMM Seeing & 0.60 to 0.75 & 0.61 to 0.70 & 0.83 to 1.32 \\
\enddata

\end{deluxetable}

NaCo is used with the L27 camera, the NB4.05 filter ($\lambda_c = 4.051
\mu m$ and $\Delta\lambda = 0.02 \mu m$) and the APP Coronagraph used to
provide additional diffraction suppression \citep{Kenworthy10a,Quanz10}.
Changes in the optical figure of the telescope mirrors and science
camera optics lead to changes in the intensity of the telescope PSF at
the science camera detector. The pupil tracking mode is therefore used
to minimise the number of optical surfaces rotating with respect to the
orientation of the detector, removing an additional component of PSF
variation. The plane of the sky therefore rotates with respect to the
camera focal plane.

Data are obtained in data cubes consisting of 200 sequential images,
each one with a duration of 0.23 seconds. One or more data cubes are
obtained in one pointing, and then the telescope is moved by 6
arcseconds along the top edge of the detector and another data cube
acquired. For these observations we use a three point dither pattern
along the top third of the NaCo detector. The APP coronagraph provides
diffraction suppression over a 180 degree wedge on one side of the star,
at a cost of brighter diffraction (and increased noise) on the other
side of the stellar PSF, resulting in a ``dark side'' and ``bright
side''. In all subsequent data reduction, the bright side of the stellar
PSF is masked off and removed from further analysis. Observations in
pupil tracking mode result in a rotation of the ``dark side'' of the PSF
on the sky.  In order to obtain high contrast sensitivity for all
position angles around the target star, at least one additional
observing block is required with the NaCo rotation set so that they
directly complement the first set of observations in position angle. In
the case for Fomalhaut we have three separate sets of APP observations
taken on separate nights (with PA offsets of $0^\circ$, $120^\circ$ and
$240^\circ$) that are combined to form a final image with 360 degrees of
sensitivity. We deliberately saturate the PSF core so that the
sensitivity to faint companions is increased.  These observations are
listed in Table \ref{tabobs}. There are two sources of loss of
sensitivity compared to direct imaging - first, the requirement to have
two complementary position angles to provide the angular coverage, and a
second loss in the core flux of all point sources in the image when
using the APP. The APP coronagraph uses 45-50\% of the core flux to
provide suppression of the diffraction rings, and this leads to an
increase of the integration time to reach a desired background limited
signal to noise limit. However, coronagraphs are always used at small
angular separations where the dominant noise source is due to
diffraction structures, and so there is a net gain in sensitivity when
using a coronagraph in these conditions.

For the photometric calibation we used unsaturated images of Fomalhaut,
itself a bright photometric standard star \citep{vanderBliek96} for the
near-IR. Based on the unpublished UKIRT bright standard star catalogue
\citep[see ][ for a discussion]{Leggett03} the UKIRT L' magnitude of
Fomalhaut is $0.96\pm 0.01$. Since $(K-L')=0\pm0.02$ for Fomalhaut and
the UKIRT L' filter is narrower than that of the more recent MKO-NIR
filter set, we take the photometry of Fomalhaut in the NB4.05 filter to
be $0.96\pm0.02$.  The observing strategy and dither positions for
Hemisphere 1 were used with the detector integration time (DIT) set to
0.0203 seconds, and 6 data cubes were obtained. We follow the general
data reduction approach in \citet{Quanz10}. With the large amount of sky
rotation in the images, we use Angular Differential Imaging
\citep{Marois06} in combination with the LOCI algorithm
\citep{Lafreniere07} to subtract the stellar PSF of our images. For each
science image, LOCI constructs a linear combination of the other science
images that have enough sky rotation to avoid faint companion
self-subtraction. The coefficients for the linear combination are
calculated for different radii from the central star and for parameters
describing the optimization region.

\section{Results and Analysis\label{resan}}

\subsection{Point source detection limits}

\begin{figure}
\centering
\includegraphics[width=.9\textwidth,angle=270]{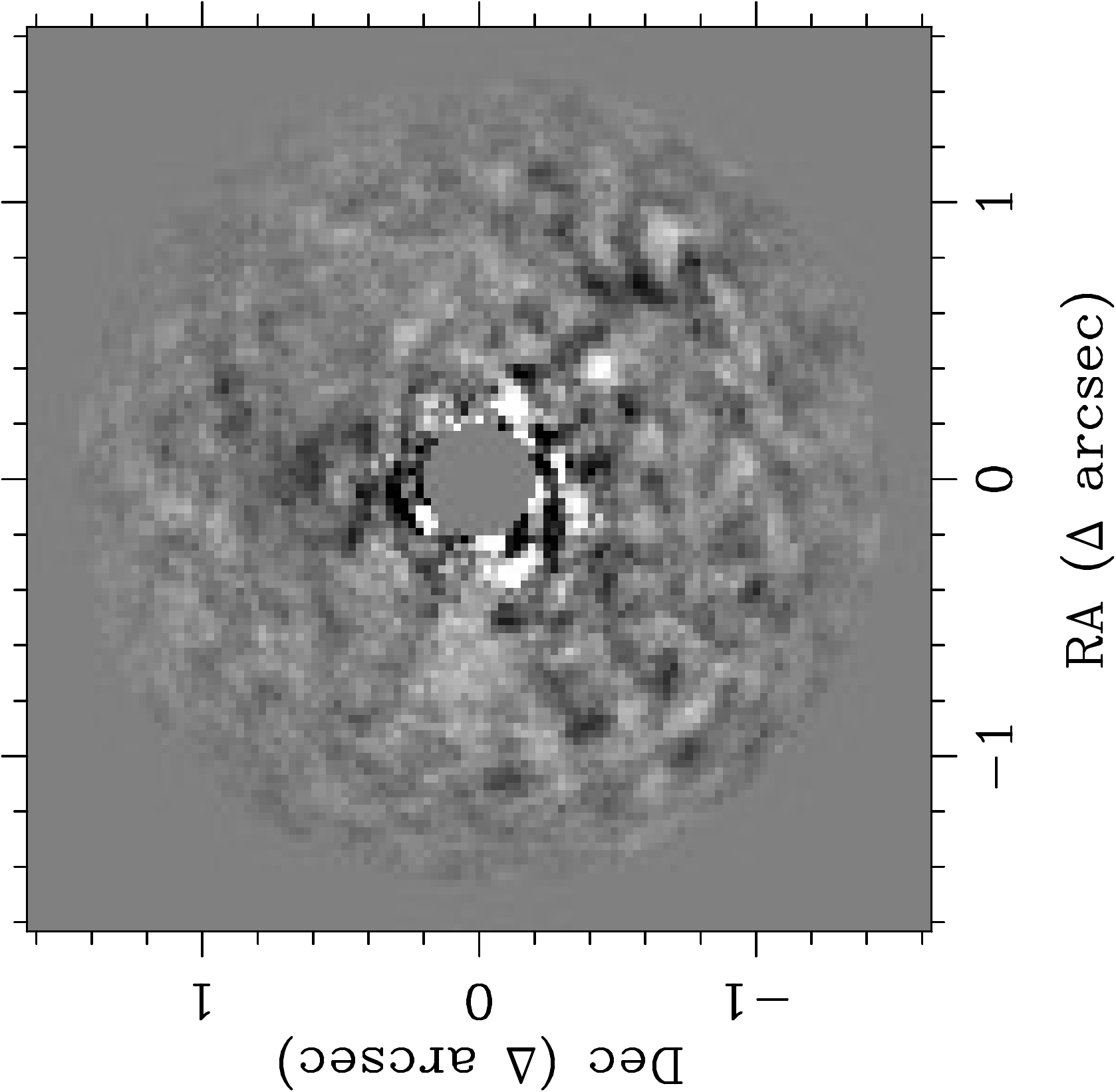}

\caption{Image of Fomalhaut with fake companion added at PA=225 and
separation of 0.6 arcseconds (4.6 AU). The companion is 11 magnitudes fainter
than Fomalhaut and is detected with a signal to noise of 8.7.
\label{fplanet}}
\end{figure}

In Figure \ref{fplanet} we show the final PSF-subtracted image for Fomalhaut,
together with an artifically inserted point source 11 magnitudes fainter
than Fomalhaut at PA=225 degrees.

\begin{figure}
\centering
\includegraphics[width=.6\textwidth, angle=270]{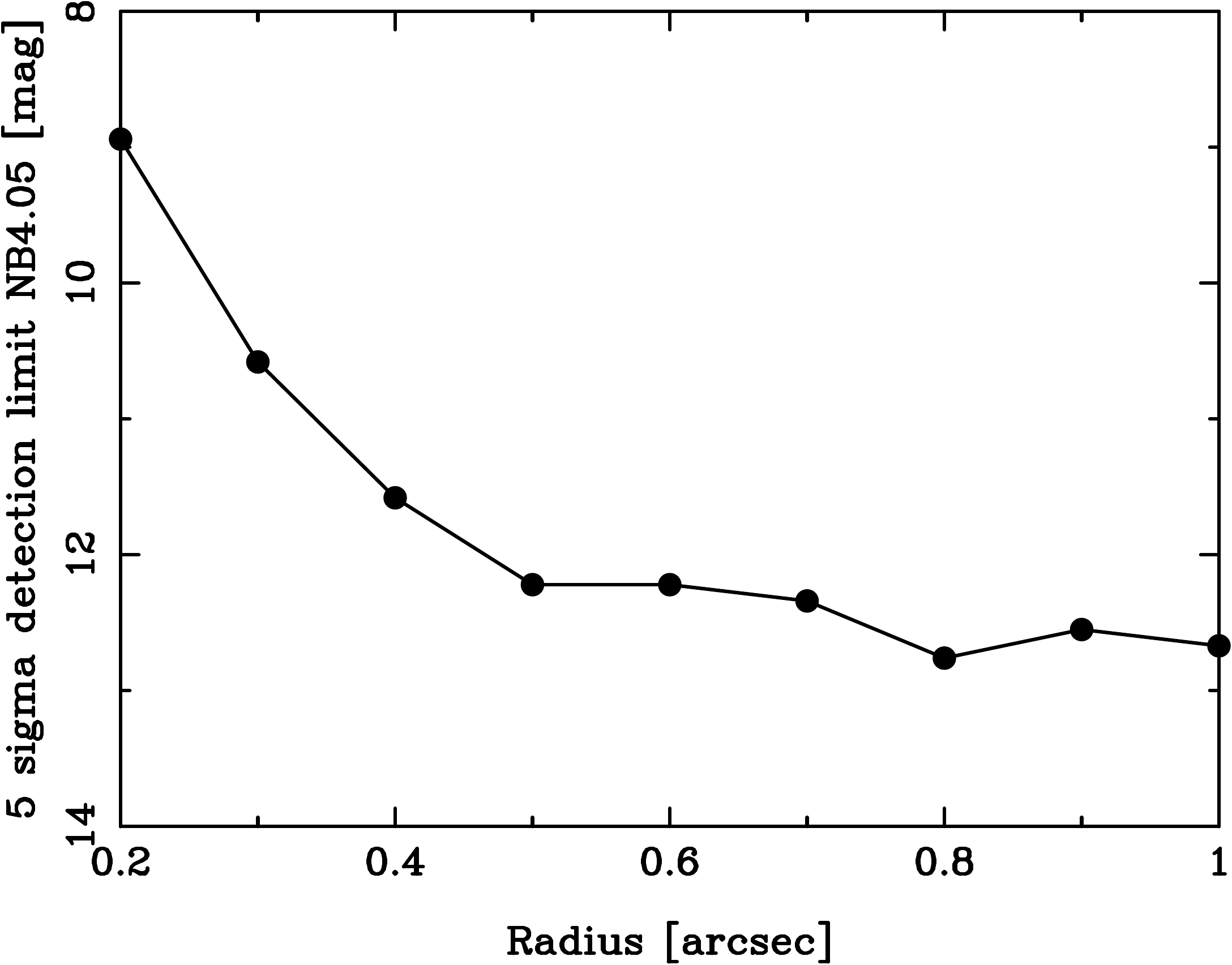}

\caption{Sensitivity curve derived from artificially inserted and
extracted point sources around Fomalhaut, with the sensitivity expressed
in apparent NB4.05 magnitudes for a $5\sigma$ point source
detection and angular separation in arcseconds. To convert to the
contrast curve, subtract 0.96 mag (see text for discussion).
\label{fomsense}}
\end{figure}

Fake planets are added to the raw data cubes to determine the
limiting magnitude achievable. The unsaturated PSF image of Fomalhaut
is used to create the fake planets in the saturated science images.
They are added at angular separations incrementing from
$0\farcs2$ to 1\arcsec\, in steps of $0\farcs1$, and delta magnitudes in
steps of 1 mag from 8 to 14.  The LOCI
algorithm is used to extract the artificial planet while minimizing
the speckles. We performed  Monte Carlo simulations to determine the
optimal parameters for the LOCI algorithm that yield the highest
signal-to-noise for the planet. The signal-to-noise for the planet is
defined as follows:

\[\left (\frac{S}{N}  \right )_{planet}=
\frac{F_{planet}}{\sigma(r)\sqrt{\pi r_{ap}^{2}}}\]

where $F_{planet}$ is the sum of the planet flux in an aperture with
radius $r_{ap}$=3 pixels and $\sigma$ is the  root mean square of the
pixels in a $170^{\circ}$ wedge at the same radius, surrounding the
star. After running the three hemispheres  through LOCI, the output was
collapsed into one final image with the best planet signal-to-noise.  We
extrapolate between the planet magnitudes to determine the magnitude
which yields a signal-to-noise of 5. This process was repeated for all
angular separations.  Fake planets were added at four different position
angles around Fomalhaut. The limiting magnitudes at each position angle
were calculated and the resultant mean limiting magnitude is shown in
Figure \ref{fomsense}.

\begin{figure}
\centering
\includegraphics[width=.7\textwidth, angle=270]{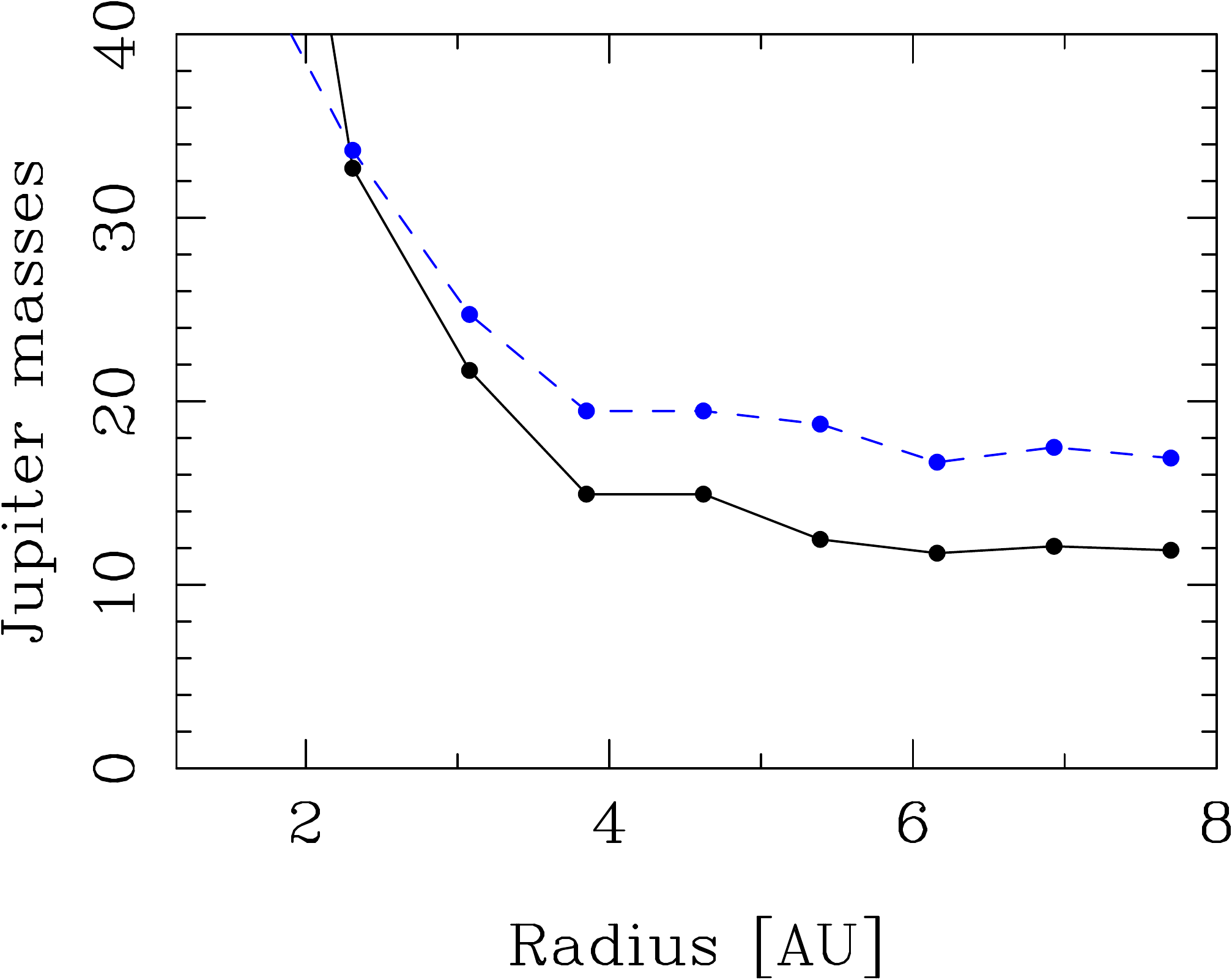}

\caption{Detection limits for gas giant companions around Fomalhaut
using \citet{Baraffe03} (solid black line) and \citet{Spiegel12} (dashed blue line) models and assuming a distance of 7.81 pc for
Fomalhaut with the best LOCI parameter case.
\label{fomplanets}}
\end{figure}

To convert a given point source's absolute magnitude to the mass of a
companion, we require both its age and a theory for the luminous evolution of the
object. We convert the contrast curve into an upper limit
of mass using the gas giant atmospheric models of \citet{Baraffe03} and
\citet{Spiegel12} to form the resultant contrast curve in Figure
\ref{fomplanets}. The \citet{Spiegel12} models are solar metallicity with
hybrid clouds. The blue curve is extrapolated from Figure 7 in
\citet{Spiegel12} to include masses greater than 10 $M_{J}$ in the
NB4.05 filter. The hot-
and cold-start models are nearly equivalent for objects older than 400
Myr, hence only one curve is shown. No significant point sources are
identified above the local azimithally averaged background out to 1.5
arcseconds (11.6 AU) around Fomalhaut. Our sensitivity in mass is flat
from 0.6 to 1.5 arcseconds (4.6 to 11.6 AU), with an upper mass limit of
$16M_{jup}$ \citep{Spiegel12} to $12M_{jup}$ \citep{Baraffe03}.

Our sensitivity to planetary mass objects is limited by the contrast
achievable using the APP optic and by the recently revised age of the
Fomalhaut system \citep{Mamajek12b}. The luminosity of giant planets
decreases as a function of time, and previous detection limits have
taken an age of Fomalhaut of $250$ Myr \citep{Navascues97}, based on
several different aging techniques on both Fomalhaut and its associated
companion TW PsA. In \citet{Mamajek12b} the companionship is further
confirmed and an improved isochronal age increases the age to $440 \pm
40$ Myr, which is the age we adopt in this paper. The consequence of the
increase in age means that previous mass sensitivities are
approximately doubled, and our results are limited down to the high end
of the planetary mass regime.

\section{Discussion\label{discus}}

\begin{figure}
\centering
\includegraphics[width=.9\textwidth,
angle=270]{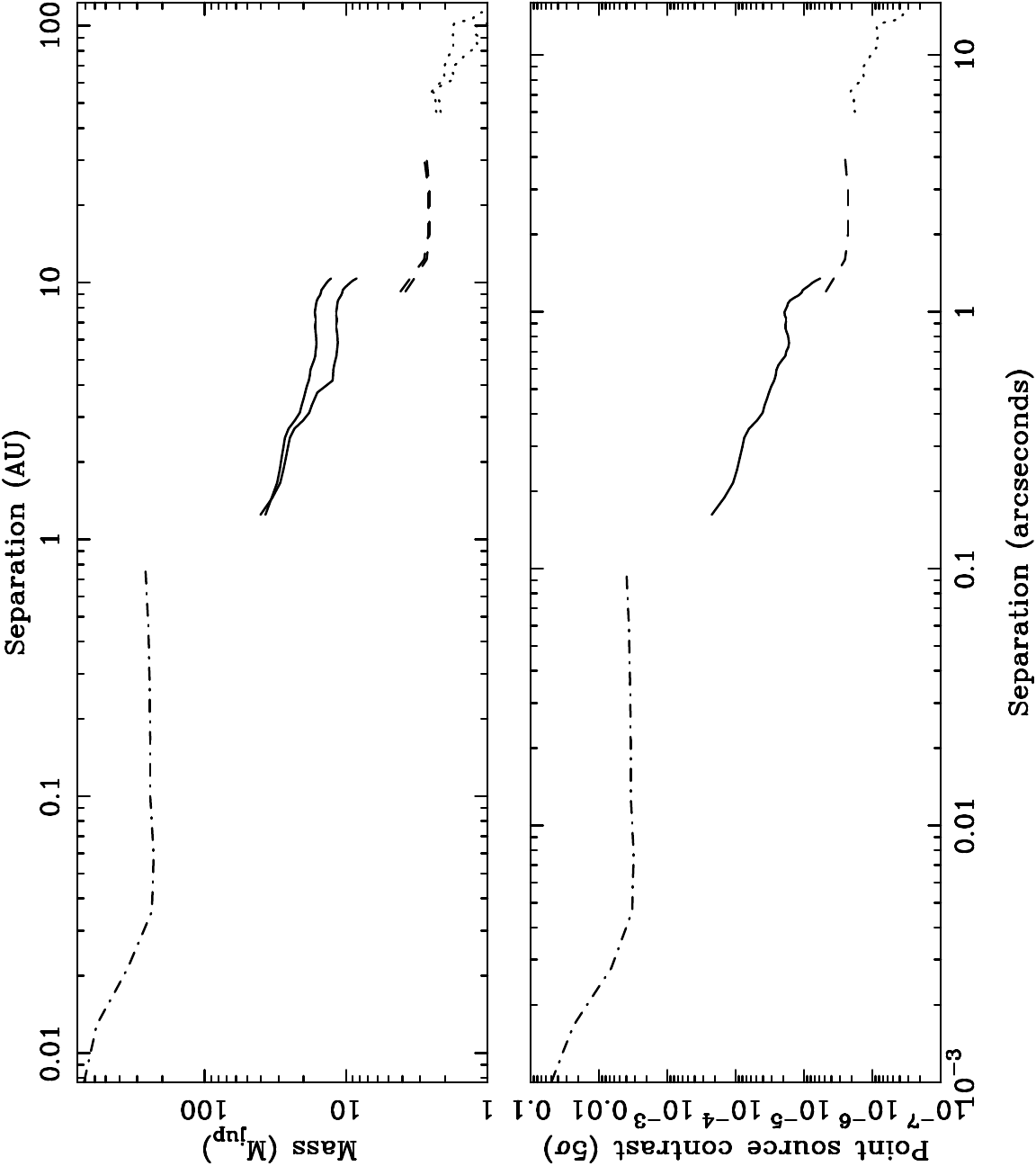}

\caption{Contrast curves and upper mass limits for point sources around
Fomalhaut compiled with measurements from the literature. From small to large angular separations, the curves
represent data from \citet{Absil11} [dot-dashed line - VLTI/1.8m/$1.65
\mu m$],
this paper [solid line - VLT/8.4m/$4.05\mu m$], \citet{Kenworthy09} [dashed
line - MMT/6.35m/$4.7\mu m$] and
\citet{Janson12} [dotted line - Spitzer/0.85m/$4.5\mu m$].
All curves have been adjusted to $5\sigma$ point source sensitivity.
Masses for the VLTI observations were estimated using the models of
\citet{Baraffe98} and the others from the models of \citet{Baraffe03} and
\citep{Spiegel12}. The age of Fomalhaut is taken to be 440 Myr \citep{Mamajek12}.
\label{allcontrasts}}
\end{figure}

Several groups have searched for massive companions in the Fomalhaut
system and these are shown in Figure \ref{allcontrasts}, both as a point
source contrast plot and an upper mass limit plot using the giant
planet models of \citet{Baraffe03} and \citep{Spiegel12}.
The VLTI/PIONIER interferometer using the auxilliary 1.8m telescopes
at H band gives the largest angular separation coverage in logarithmic
space from 1mas to 100mas with a peak sensitivity of $3\times10^{-3}$,
scaled for a point source contrast limit at $5\sigma$ \citep{Absil11}.
This angular separation regime is not accessible to single dish
telescopes, and future observations with the full 8.4m telescopes in
the VLTI will be able to look within 1 AU of Fomalhaut and other nearby
bright stars.

This work searches from 0.2 to 1.2 arcseconds using a coronagraphic
optic that provides additional diffraction suppression and provides and
additional 1 to 1.5 magnitudes of sensitivity compared to direct imaging
sensitivities obtained with the same wavelength and instrument
\citep{Quanz12}.
These observations overlap with the thermal infrared ($4.7\mu m$)
measurements from \citet{Kenworthy09} to look for companions out to 35
AU, where the planet detection limit reaches down to $2.6M_{jup}$ (using
the models of \citet{Baraffe03} and the revised age of 440 Myr), taking
advantage of the thermal imaging performance of the 6.5m MMTO telescope
\citep{LloydHart00} with the Clio thermal infrared imager \citep{Hinz06}
and deformable secondary AO mirror system \citep{Wildi02,Brusa03}.  The
observations in \citet{Kenworthy09} are complete out to 4.5 arcseconds,
and additionally provide partial coverage to 10 arcseconds. We
restrict the contrast curve in Figure \ref{allcontrasts} to the complete
coverage.

Others have looked for the thermal signature of Fomalhaut b using
the Spitzer space telescope \citep{Marengo09,Janson12} with model
dependent upper mass limits of $1M_{jup}$ \citep{Janson12} being reached
with the lower background sensitivity achievable in space. This also
provides a constraint on companions in to 11 arcseconds.

The contrasts attained at 1.3 arcseconds between the
MMT and VLT observations are comparable while there is a factor of two in
planet mass sensitivity achieved. This is due to the extremely red
colours predicted (and seen) in extrasolar giant planet models for
younger systems $(< 500 Myr)$, demonstrating the utility of carrying out
extrasolar planet searches at thermal wavelengths around nearby stars
\citep{Hinz06,Heinze10b} with both current facilities
\citep[LMIRCam][]{Wilson08} and future ELTs such as the E-ELT
\citep[METIS;][]{Brandl12} and GMT \citep[TIGER;][]{Hinz12}. By
optimizing the telescope for thermal infrared observations with a
deformable secondary mirror \citep{Brusa03a}, fainter planets can be
detected \citep{LloydHart00}.

There is a gap in coverage from 100 mas to 200 mas, which can be
bridged with Sparse Aperture Masking (SAM) techniques
\citep{Nakajima89,Tuthill06,Tuthill00} to cover 100 to 500
mas. This technique is well suited to bright targets such as Fomalhaut,
and it is expected that this semi-major axis region will be observed in
due course.

As can be seen, there is a large degree of coverage in semi-major axis
around Fomalhaut, ruling out the presence of massive planetary mass companions
to within approximately 3 AU and brown dwarf companions within 2 AU.
Further epochs will also close out the phase space that remains open due
to the non-simultaneity of the reported observations.

\citet{Crida09} suggest a pair of massive planets at 5-20 AU could
undergo a change in their semi-major axes through a mean motion
resonance, with the larger companion remaining at 75 AU whilst the outer
one moves out to form the core for Fomalhaut b.  The search presented
here, in combination with other ground and space-based observations
rules out the presence of brown dwarf mass companions. The need
for a large, inner system scattering planet, however, may not be
required to form Fomalhaut b out at the large distance from the central
star that it currently appears. \citet{Lambrechts12} suggest that gas
giant cores can be rapidly accreted at large stellar distances in the
early gas-rich phase of the protoplanetary disk by pebble accretion,
obviating the need for gravitational scattering from the inner stellar
system.

\section{Summary and Conclusions\label{concl}}

We have presented upper limits to massive companions in the Fomalhaut
system, covering the range of semi-major axes from 1.5 AU to 11 AU. We
detect no companions greater than a model dependent $20 M_{jup}$ from 4
AU to 10 AU, and combining this with other searches in the literature we
can rule any companion greater than $30M_{jup}$ from 2.5 AU outwards.
The Fomalhaut system is older than previously thought, leading to an
upward revision of the previously determined masses. We also rule out
any brown dwarf that can cause the anomalous astrometric motion of
Fomalhaut in Hipparcos data as suggested by \citet{Chiang09}. In the
case of Fomalhaut, SAM observations are
expected to close the gap for 0.1 to 0.5 arcseconds.

Future research is needed to confirm that the contrast limits at small
angular separations are now dominated by non-common path errors between
the wavefront sensor optical path and the science camera optical path.
Efforts to measure and remove these errors have been demonstrated on NaCo with
extra-focal imaging \citep{Riaud12} and by using the residuals from the
wavefront sensor camera in a closed loop AO system \citep[][and Codona
and Kenworthy; submitted]{Codona08}.

\acknowledgments

This research has made use of the SIMBAD database, operated at CDS,
Strasbourg, France.  We are indebted to the ESO Paranal Support staff at
the VLT, and to the anonymous referee for their comments on this
paper.  The data here were obtained under ESO Observing Run Number
087.C-0701(B).

{\it Facilities:} \facility{VLT}.

\clearpage

\bibliographystyle{nsf}
\bibliography{apj-jour,kenworthy}

\end{document}